\documentclass[12pt,preprint]{aastex}

\shorttitle{Light Elements and $r$-Process Elements in SNe}
\shortauthors{Yoshida et al.}

\begin{document}


\title{Nucleosynthesis of Light Elements and Heavy $r$-Process Elements 
through the $\nu$-Process in Supernova Explosions}


\author{Takashi Yoshida\altaffilmark{1,2}, Mariko Terasawa\altaffilmark{3,4}, 
Toshitaka Kajino\altaffilmark{3,5}, and Kohsuke Sumiyoshi\altaffilmark{6}}
\affil{$^1$Department of Physics, School of Sciences, Kyushu University, 
Ropponmatsu, Fukuoka 810-8560, Japan}
\affil{$^3$National Astronomical Observatory, and The Graduate 
University for Advanced Studies, 2-21-1 Osawa, Mitaka, Tokyo 181-8588, Japan}
\affil{$^5$Advanced Science Research Center, Japan Atomic Energy 
Research Institute, Tokai, Naka, Ibaraki 319-1195, Japan}
\affil{$^6$Numazu College of Technology, Ooka, Numazu, 
Shizuoka 410-8501, Japan}

%\and


\altaffiltext{2}{Current address: National Astronomical Observatory,
2-21-1 Osawa, Mitaka, Tokyo 181-8588, Japan; takashi.yoshida@nao.ac.jp}
\altaffiltext{4}{Current address: Center for Nuclear Study, 
Graduate School of Science, University of Tokyo, Hirosawa, Wako, 
Saitama 351-0198, Japan}


\begin{abstract}
We study the nucleosynthesis of the light elements $^7$Li and $^{11}$B and 
the $r$-process elements in Type II supernovae from the point of view 
of supernova neutrinos and Galactic chemical evolution.
We investigate the influence of the luminosity and average energy 
(temperature) of supernova neutrinos on these two nucleosynthesis processes.
Common models of the neutrino luminosity, which is parameterized by 
the total energy $E_{\nu}$ and decay time $\tau_{\nu}$ and neutrino 
temperature are adopted to understand both processes.
We adopt the model of the supernova explosion of a 16.2 $M_{\odot}$ star, 
which corresponds to SN 1987A, and calculate the nucleosynthesis of the light 
elements by postprocessing.  
We find that the ejected masses of $^7$Li and $^{11}$B are
roughly proportional to the total neutrino energy and are
weakly dependent on the decay time of the neutrino luminosity.
As for the $r$-process nucleosynthesis, we adopt the same models of 
the neutrino luminosity in the neutrino-driven wind models of 
a 1.4 $M_{\odot}$ neutron star.
We find that the $r$-process nucleosynthesis is affected through the peak 
neutrino luminosity, which depends on $E_{\nu}/\tau_{\nu}$.
The observed $r$-process abundance pattern is better reproduced at a
low peak neutrino luminosity.
We also discuss the unresolved problem of the overproduction of $^{11}$B
in the Galactic chemical evolution of the light elements.
We first identify that the ejected mass of $^{11}$B is a factor of 2.5-5.5 
overproduced in Type II supernovae when one adopts neutrino 
parameters similar to those in previous studies, 
i.e., $E_{\nu} = 3.0 \times 10^{53}$ ergs, $\tau_{\nu} = 3$ s, and a neutrino 
temperature 
$T_{\nu_{\mu, \tau}} = T_{\bar{\nu}_{\mu, \tau}} = 8.0$ MeV/$k$.
We have to assume $E_{\nu} \le 1.2 \times 10^{53}$ ergs to avoid
the overproduction of $^{11}$B, which is too small to accept in comparison 
to the $3.0 \times 10^{53}$ ergs deduced from the observation of SN1987A. 
We here propose to reduce the temperatures of $\nu_{\mu, \tau}$ and 
$\bar{\nu}_{\mu, \tau}$ to 6.0 MeV/$k$ in a model with 
$E_{\nu} \sim 3.0 \times 10^{53}$ ergs and $\tau_{\nu} \sim 9$ s.
This modification of the neutrino temperature is shown to resolve 
the overproduction problem of $^{11}$B while still keeping a successful 
$r$-process abundance pattern.
\end{abstract} 




\section{Introduction}

During supernova explosions, a huge amount of neutrinos are emitted, 
blowing off surface materials from the proto-neutron star.
The neutrinos interact with nuclei in the supernova ejecta, and the neutrino 
emission is strong enough to change the compositions, despite the small 
cross sections for neutrino-nucleus interactions.
The neutrino-induced reactions mainly affect two kinds of proposed
nucleosynthetic processes that occur during supernova explosions: one is
the synthesis of light elements such as Li and B through the $\nu$-process 
in the He-layer, and the other is the $r$-process in the neutrino-driven 
winds above the surface of the neutron star.

The production of light elements through the $\nu$-process during 
supernova explosions was first suggested by Domogatsky, Eramzhyan, \& 
Nadyozhin (1977).
Woosley et al. (1990) precisely evaluated the roles of the $\nu$-process
and showed that a large amount of $^7$Li and $^{11}$B is produced during
supernova explosions.
Woosley \& Weaver (1995, hereafter WW95) tabulated the abundances of 
the elements, including the light elements with grids of stellar masses 
and metallicities. 
Their results have been adopted for studies on Galactic chemical 
evolution (GCE; e.g., Fields et al. 2000; Ramaty et al. 2000b; 
Ryan et al. 2001).

However, Olive et al. (1994) pointed out that there remains a serious problem 
of overproduction of $^{11}$B from the supernova $\nu$-process in the GCE models of 
the light elements.
Here the overproduction means that the predicted $^{11}$B abundance in 
theoretical calculations is overabundant compared to the observed one when 
we adopt the theoretical yields of WW95 without any renormalization.
Studies on GCE have shown that light elements are mostly produced 
from Galactic cosmic-ray (GCR) interactions with the interstellar medium 
(ISM).
The GCR model was improved by taking account of the primary acceleration of 
heavy elements from supernova ejecta in addition to the secondary 
acceleration of the engulfing ISM (Ramaty et al. 1997; 
Yoshii, Kajino, \& Ryan 1997; Vangioni-Flam et al. 1998; 
Suzuki, Yoshii, \& Kajino 1999).
This explains the naturally linear metal dependence of the amounts of Be and B 
during GCE; i.e., [BeB/H] $\propto$ [Fe/H], but one still needs another 
contribution to $^7$Li and $^{11}$B (Olive et al. 1994; Vangioni-Flam et al. 
1996; Fields \& Olive 1999; Romano et al. 1999).
The production of $^{11}$B in the supernova $\nu$-process is identified as 
the most important process for explaining the very precise data of the 
meteoritic $^{11}$B/$^{10}$B abundance ratio. 
Several authors studied the GCE models of the light elements by taking account 
of the contribution of the supernova $\nu$-process.
They showed that the amount of $^{11}$B is too large by a factor of 2 
(Fields et al. 2000) to 5 (Ramaty, Lingenfelter, \& Kozolvsky 2000a), 
while the other light elements $^6$Li, $^7$Li, $^9$Be, and $^{10}$B are 
well reproduced in the appropriate amount.

Mass loss of the outer envelope in the presupernova evolutionary phase would 
decrease the efficiency of $^{11}$B production in the $\nu$-process during the 
supernova explosion.
Wolf-Rayet stars in fact exhibit strong activities in their stellar 
atmosphere, such as mass loss.
They originate from stars as massive as 40 $M_{\odot}$ (e.g., 
Abbott \& Conti 1987; Meynet et al. 2001).
Since the supernovae discussed in the present article have main-sequence 
masses of 13-30 $M_{\odot}$, there is no need to take account 
of such a mass-loss effect.

The production of $^7$Li and $^{11}$B during the supernova explosion depends 
on the supernova models.
It also depends on the details of the total neutrino luminosity and its time 
variation.
Moreover, the yet uncertain average neutrino energy should strongly affect the 
$\nu$-spallation cross sections of $^4$He, providing seed elements for the 
production of $^7$Li and $^{11}$B (WW95).
Since the supernova neutrinos would therefore affect the final abundance of 
$^7$Li and $^{11}$B (Fields et al. 2000; Yoshida, Emori, \& Nakazawa 2000), 
we should investigate the dependence on the 
neutrino spectra to solve the overproduction problem.

Following the core explosion of a supernova, a ^^ ^^ hot bubble'' region, 
in which the density is relatively low and the temperature and entropy are 
high, is formed between the surface of the proto-neutron star and the outward 
shock wave.
In the region near the surface of the proto-neutron star, the material is 
blown off by neutrino heating.
The outflow of the material is also called the ^^ ^^ neutrino-driven wind''.
Woosley et al. (1994) showed that the $r$-process occurs successfully 
in neutrino-driven winds with very high entropy, $400k$, where $k$ denotes 
the Boltzmann constant.
However, their nucleosynthesis calculation did not include 
neutrino-nucleus interactions during the postprocessing of the nucleosynthesis.
It was subsequently pointed out that the supernova neutrinos convert neutrons 
into protons during nucleosynthesis and that the $r$-process has difficulty 
in producing third-peak elements because of the neutron deficiency even 
in a high-entropy hot bubble (Fuller \& Meyer 1995; Meyer 1995).
It was also reported that independent simulations of neutrino-driven winds 
have difficulty in producing the required high-entropy condition 
(Witti, Janka \& Takahashi 1994; Takahashi, Witti, \& Janka 1994).
Thus, neutrino-driven winds were suspected to be a site of the $r$-process.

Recently, neutrino-driven wind models have been revived as promising sites 
of $r$-process nucleosynthesis by using a moderately high entropy, 
$\sim 200 k$, and a very short expansion timescale, $\sim 10$ ms 
(Otsuki et al. 2000; Sumiyoshi et al. 2000).
Massive ($\sim 2.0 M_{\odot}$) and compact ($10$ km) neutron star models 
are assumed in order to obtain such conditions.
It is known, however, that the typical mass of a neutron star is about 
$1.4 M_{\odot}$ and that the radius is about 10 km.
Terasawa et al. (2002) have recently shown the possibility for 
a successful $r$-process abundance pattern to emerge from a neutron 
star model with a typical mass of $1.4 M_{\odot}$ and a radius of $10$ km; 
they use a slightly low asymptotic temperature at the outer boundary 
of the neutrino-driven winds.
In all these simulations, they set the mean energies of neutrinos to be about 
$10, 20$, and $30$ MeV  for $\nu_e, \bar{\nu_e}$, and $\nu_i$ 
($i = \mu, \tau$, and their antiparticles), respectively, to match with those 
adopted in a previous theoretical study (Qian \& Woosley 1996).

In light of the successful $r$-process nucleosynthesis in neutrino-driven 
winds, it is of current interest and importance to study how to solve the 
overproduction problem of the light elements in the $\nu$-process in the 
same supernova model.
Since neutrinos are very weakly interacting particles, their energy spectrum 
would not change in the ejecta unless neutrino oscillation were considered.
Nevertheless, there was no attention given to the fact that the supernova 
neutrino model, which is used for light-element synthesis during supernova 
explosions, should be identical to that adopted in the neutrino-driven 
wind models.

In the present study we use a common neutrino luminosity that
decreases with time in the application to both light-element synthesis in the 
He layer and $r$-process synthesis in neutrino-driven winds.
We investigate the sensitivity of light-element synthesis in supernova ejecta 
to the neutrino luminosity with the two parameters of the decay time 
$\tau_{\nu}$ and the total neutrino energy $E_{\nu}$.
At the same time, we simulate the neutrino-driven winds with the same 
neutrino luminosity parameters and calculate the $r$-process abundance pattern.
We thus discuss the consistency between the light-element production and the 
abundance distribution of the $r$-process elements and try to solve the 
overproduction problem of the light elements in GCE.

In addition to the ambiguity of the neutrino luminosity, neutrino temperature 
is still a controversial problem.
Although extensive studies by supernova simulations with neutrino transfer 
have been done by several groups (e.g., Janka, Kilfonidis, \& Rampp 2001; 
Liebend\"orfer et al. 2001; Thompson, Burrows, \& Pinto 2003; 
Buras et al. 2003), the explosion mechanism has not yet been clarified.
Accordingly, the information on supernova neutrinos has not been uniquely 
determined.
Detailed studies on supernova neutrinos have been made to determine the 
neutrino luminosity and spectra (Myra \& Burrows 1990; Suzuki 1994; 
Totani et al. 1998; Keil, Raffelt, \& Janka 2003).
Hence, it is also important to investigate the sensitivity of 
the light-element production and the $r$-process abundance pattern 
to the temperature of the $e$-, $\mu$-, and $\tau$-neutrino families 
using a common luminosity for the supernova neutrinos.
Recent theoretical studies on explosive nucleosynthesis in supernovae
(e.g., Rauscher et al. 2002) have shown a smaller ejected mass of $^{11}$B, 
moving toward a solution of the overproduction problem.
They assumed slightly lower temperatures of $\mu$- and $\tau$-type 
neutrinos and their antiparticles ($\nu_{\mu,\tau}$ and 
$\bar{\nu}_{\mu,\tau}$) than those adopted in WW95.
We therefore explore many different neutrino luminosities with different 
neutrino temperatures to look for an appropriate ejected mass of $^{11}$B 
and $r$-process abundance pattern.
This result would in turn strongly constrain models of supernova neutrinos.

\section{Calculations}

\subsection{Neutrino Luminosity and Temperature}

In order to investigate the relation between the ejected mass of the light
elements and the $r$-process abundance pattern, we use a common model of 
neutrino luminosity based on Woosley et al. (1990) and WW95. 
The neutrino luminosity $L_{\nu _i}$ ($\nu_i = \nu_e, \nu_{\mu}, \nu_{\tau}$, 
and their antiparticles) is the same for all species and exponentially 
decreases with a decay time $\tau_{\nu}$
\begin{equation}
L_{\nu _i}(t) = \frac{1}{6} \frac{E_{\nu}}{\tau_{\nu}} \exp \left(
-\frac{t-r/c}{\tau_{\nu}} \right) \Theta(t-r/c) \, ,
\end{equation}
where $E_{\nu}$ is the total neutrino energy, $r$ is the radius, $c$ is the 
speed of light, and $\Theta(x)$ is a step function defined by 
$\Theta(x)=1$ for $x>0$ and 0 otherwise.
The total neutrino energy and the decay time of the neutrino 
luminosity are parameters.
In the above two papers, the authors fixed $\tau_{\nu} = 3$ s and 
$E_{\nu} = 3 \times 10^{53}$ ergs. 
Here, we adopt a wider range for these parameters.
We set the decay time in the range between 1 and 3 s.
The total neutrino energy $E_{\nu}$ is evaluated as approximately 6 times 
the total energy emitted from electron antineutrinos.
The determined total energy of $\bar{\nu}_{\rm e}$ from SN 1987A (Hirata 
et al. 1987; Bionta et al. 1987) leads to a total neutrino energy ranging 
over $1.4 \times 10^{53}$ ergs $\la$ $E_{\nu}$ $\la$ 
$6.2 \times 10^{53}$ ergs (Suzuki 1994).
Since the error bars are very large, depending on the different methods 
of maximum likelihood analysis of the observed data, we vary the total 
neutrino energy $E_{\nu}$ in the range between $1.0 \times 10^{53}$ and 
$6.0 \times 10^{53}$ ergs.

When we evaluate the reaction rates of the $\nu$-process and thermal
evolution of the neutrino-driven winds, we further need to know the
energy spectra of all species of neutrinos.
Strictly speaking, the neutrino energy spectra do not have a thermal 
distribution because of the strong energy dependence of weak interactions.
However, for our present purpose of studying the $\nu$-process, 
the spectra of the neutrinos emitted from the warm surface of proto-neutron 
stars can be approximately expressed as a thermal distribution 
at a certain temperature.
This is because only the high-energy tails of the spectra are important 
for the $\nu$-process.
Therefore, both the average energy of the neutrinos and the average cross 
sections can be calculated accordingly and have been widely utilized in the 
previous nucleosynthesis calculations in a standard manner.
Although some studies showed that these neutrino spectra may have a non zero 
chemical potential (e.g., Myra \& Burrows 1990; Hartmann et al. 1999; 
Keil et al. 2003), we assume that the chemical potential of the neutrinos is 
zero for all species, as done in the previous studies of light-element 
nucleosynthesis (Woosley et al. 1990; WW95; Rauscher et al. 2002).

We set the temperature of $\nu_{\mu,\tau}$ and $\bar{\nu}_{\mu,\tau}$ as
\begin{equation}
T_{\nu _{\mu,\tau}}=T_{\bar{\nu}_{\mu,\tau}}= \frac{8.0 {\rm MeV}}{k}.
\end{equation}
This value is taken from Woosley et al. (1990) and WW95.
The temperature corresponds to a mean energy of 25 MeV.
The temperatures of $\nu_{\rm e}$ and $\bar{\nu}_{\rm e}$ are set to be 
\begin{equation}
T_{\nu_{\rm e}} = \frac{3.2 \  {\rm MeV}}{k} ,
\end{equation}
\begin{equation}
T_{\bar{\nu}_{\rm e}} = \frac{5.0 \  {\rm MeV}}{k} ,
\end{equation} 
respectively.
The corresponding mean energies are 10 and 16 MeV.

Let us remark that the temperatures of $\nu_{\rm e}$ and $\bar{\nu}_{\rm e}$ 
are different from those in WW95, who assumed the same temperature of 
4 MeV/$k$ for both $\nu_{\rm e}$ and $\bar{\nu}_{\rm e}$.
More detailed numerical studies of supernova neutrinos (e.g., Woosley 
et al. 1994; Janka et al. 2001) showed 
that $T_{\nu_{\rm e}}$ is lower than $T_{\bar{\nu}_{\rm e}}$ because 
the interactions of $\bar{\nu}_{\rm e}$ with protons freeze out at a higher 
density and temperature, whereas $\nu_{\rm e}$ interacts with 
neutrons also at a lower density and temperature.
We also note that a temperature for $\nu_{\mu,\tau}$ and 
$\bar{\nu}_{\mu,\tau}$ lower than 8 MeV/$k$ has been reported in recent
studies of supernova neutrino spectra formation (Myra \& Burrows 1990; 
Keil et al. 2003).
A temperature lower than that in equation (2), i.e., 
$T_{\nu_{\mu,\tau}}=T_{\bar{\nu}_{\mu,\tau}}$= 6 MeV/$k$ 
, is also used in discussions below, as well as 8 MeV/$k$.

\subsection{Supernova Explosion Model for Light Element Nucleosynthesis}

We study light-element synthesis by a postprocessing nucleosynthesis
calculation of a supernova explosion.
The presupernova model is the 14E1 model (Shigeyama \& Nomoto 1990), 
corresponding to a model for SN 1987A. 
This model is constructed from a precollapse 6 $M_{\odot}$ helium star 
(Nomoto \& Hashimoto 1988) and a 10.2 $M_{\odot}$ H-rich envelope.
The chemical composition of the presupernova model is taken from the 14E1 
model.
In the H-rich envelope, the mass fractions of $^1$H and $^4$He are set to
be $X=0.565$ and $Y=0.43$, respectively.
For the postprocessing calculation, the abundance distribution of the
CNO-elements is assumed to be the equilibrium values of the CNO-cycle.
The abundances of heavier elements are assumed to be one-third of those of the
sola$r$-system abundances, i.e., $Z=0.005$.

In order to calculate the light-element synthesis in the supernova
by postprocessing, we have to know the time evolutions of the temperature, 
density, and radius during the supernova explosion.
In the present study we evaluate the propagation of a shock wave during the
supernova explosion using a spherically symmetric Lagrangian PPM (piecewise 
parabolic method) code (Colella \& Woodward 1984; Shigeyama et al. 1992) 
which includes a small nuclear reaction network containing 13 
kinds of $\alpha$-nuclei.
The explosion energy is set to be $1 \times 10^{51}$ ergs and the
location of the mass cut is assumed to be 1.61 $M_{\odot}$.

A nuclear reaction network for the postprocessing calculation of the light
element synthesis consists of 291 nuclear species up to Ge (Table 1).
Reaction rates in this network are adopted from NACRE (Angulo et al. 1999), 
Caughlan \& Fowler (1988), Rauscher \& Thielemann (2000),
Bao et al. (2000), Horiguchi et al. (1996), 
Fuller, Fowler, \& Newman (1982), and Oda et al. (1994).
The reaction rates of the $\nu$-process are adopted from the 1992 work by 
R. D. Hoffman \& S. E. Woosley\footnote[7]{See 
http://www-phys.llnl.gov/Research/RRSN/nu\_csbr/neu\_rate.html.}.

\subsection{Neutrino-Driven Wind Models for the $r$-Process Nucleosynthesis}

We adopt models of neutrino-driven winds for the $r$-process nucleosynthesis 
in a core-collapse supernova explosion. 
The thermodynamic conditions of the neutrino-driven winds strongly depend on 
the neutrino properties. 
We use the same neutrino luminosity and temperature as we described 
in section 2.1 for numerical simulations of the neutrino-driven winds 
and the $r$-process nucleosynthesis.
Instead of following the time evolution of the neutrino luminosity, we run a 
set of different simulations of the neutrino-driven winds with constant and 
different neutrino luminosities, and superpose the calculated results of 
nucleosynthesis as employed by Wanajo et al. (2001, 2002).
This is a good approximation because the decay time of the neutrino 
luminosity with $\tau_{\nu}$ = 1-3 s is long enough compared to the 
expansion timescale of the neutrino-driven winds and the nuclear reaction 
timescale of $\alpha$- and $r$-processes 
(Otsuki et al. 2000; Terasawa et al. 2001).
We refer to the time $t = t_{end}$ as when the luminosity decays to as low as 
$L_{\nu_i} = 3.5 \times 10^{51}$ ergs s$^{-1}$. 
We choose three representative times at $0, t_{end}/2$, and $t_{end}$ 
for each model of neutrino luminosity and superpose the three calculated  
results of hydrodynamic simulations and $r$-process nucleosynthesis 
(see Table \ref{table1}).

To follow the hydrodynamic evolution of neutrino-driven winds, we employ an 
implicit numerical code for general-relativistic and spherically symmetric 
hydrodynamics (Yamada 1997; Sumiyoshi et al. 2000), including the heating 
and cooling processes due to neutrinos (Qian \& Woosley 1996).
As an initial condition, we put thin surface material on a neutron
star with a typical mass of $1.4 M_{\odot}$ and radius of $10$ km, 
which is the inner boundary condition. 
We obtain the initial structure of this material by solving 
the Oppenheimer-Volkoff equation. 
As an outer boundary condition, we put a constant pressure $P_{out}$ next
to the outermost grid point of the Lagrangian mesh. The value of $P_{out}$ is 
taken to be $10^{20}$ dyn cm$^{-2}$ for all simulations, since 
a low asymptotic temperature due to a low outer pressure is favorable for 
$r$-process nucleosynthesis (Terasawa et al. 2002).

We start the network calculations of the
nucleosynthesis at the time when the temperature drops to $T_9 = 9.0$ (in
units of $10^9$ K) and follow the time evolution of the abundances.
The reaction network covers over 3000 species of nuclei from the
$\beta$-stability line to the neutron drip line including light
neutron-rich unstable nuclei (Terasawa et al. 2001). 
It includes only the charged-current neutrino-nucleus interactions for 
all nuclei. Since the neutral-current interactions have little influence on 
the final composition of the material when the timescale of expansion is very 
short (Terasawa et al. 2003), as in our present model, 
we did not include these reactions in the current studies of the $r$-process.

In order to estimate the total ejected mass of each isotope, $M_{eject,i}$,
for a given neutrino luminosity, we sum up the three results of 
nucleosynthesis ensembles with different neutrino luminosities in 
the following trapezoid formula:
\begin{equation}
M_{{\rm eject},i} =  \left( \frac{\dot M_{0,i} + \dot M_{{\rm half},i}}{2}
 + \frac{\dot M_{{\rm half},i} + \dot M_{{\rm end},i}}{2} \right)
\frac{t_{{\rm end}}}{2} ,
\label{superpose}
\end{equation}
where $t_{end}$ is the time when the luminosity becomes 
$3.5 \times 10^{51}$ ergs s$^{-1}$, as defined before, and 
$\dot M_{0,i}, \dot M_{{\rm half},i}$, and $\dot M_{{\rm end},i}$ 
are the mass ejection rates of isotope $i$ obtained from 
the calculation for the three neutrino luminosities 
$ L_{\nu_i} = L_{\nu_i,0}, L_{\nu_i,{\rm half}}$, and $L_{\nu_i,{\rm end}}$ at 
$t = 0, t_{{\rm end}}/2$, and $t_{{\rm end}}$, respectively.

\section{Results}
\subsection{Abundances of $^7$Li and $^{11}$B}
We examine the influence of the neutrino luminosity on 
the ejected masses of $^7$Li and $^{11}$B.
Figure 1 shows the mass fractions of the light elements as a function of
the mass coordinate $M_r$ in the case of $E_{\nu}=3 \times 10^{53}$ ergs 
and $\tau_{\nu}=3$ s, which is the same parameter set as adopted in WW95.
We see that $^7$Li and $^{11}$B are abundantly produced in the He/C layer 
in the ranges of 
$4.6 M_{\odot} \lesssim M_r \lesssim 5.8 M_{\odot}$ and
$4.2 M_{\odot} \lesssim M_r \lesssim 5.0 M_{\odot}$, respectively.
In addition, $^{10}$B is produced in the regions below the He/N layer but 
less abundantly.
A small amount of $^6$Li and $^9$Be are also produced in the O/C layer and the
outer part of the He/C layer.

Let us first explain the main production process of the light elements.
In the $^7$Li-production region 
($4.6 M_{\odot} \lesssim M_r \lesssim 5.8 M_{\odot}$), neutrinos emitted 
from the collapsed core break up 
$^4$He through $^4$He($\nu,\nu'p)^3$H and $^4$He($\nu,\nu'n)^3$He.
The produced $^3$H and $^3$He capture $^4$He to produce $^7$Li through 
$^3$H($\alpha,\gamma)^7$Li and
$^3$He($\alpha,\gamma)^7$Be(e$^-,\nu_{\rm e})^7$Li.
The main production process of $^{11}$B in the $^{11}$B production region
($4.2 M_{\odot} \lesssim M_r \lesssim 5.0 M_{\odot}$) needs another step in 
addition to the $^7$Li production; 
the produced $^7$Li leads to $^{11}$B through $^7$Li($\alpha,\gamma)^{11}$B.
Note that the $\alpha$-capture reaction from $^7$Be, i.e.,
$^7$Be($\alpha,\gamma)^{11}$C(e$^{+}\nu_{\rm e})^{11}$B, also proceeds, but
only a small amount of $^{11}$B is produced through this reaction sequence.

On both sides of the inner and outer mass coordinates surrounding 
the $^7$Li- and $^{11}$B-production regions mentioned above, the mass 
fractions of $^7$Li and $^{11}$B are smaller.
In the region below the He/C layer, the mass fraction of $^7$Li is much
smaller than in the He/C layer because the mass fraction of $^4$He,
which is the seed nucleus of $^7$Li, is very small.
The mass fraction of $^{11}$B is also smaller.
In this region $^{11}$B is produced through $^{12}$C($\nu,\nu'p)^{11}$B and
$^{12}$C($\nu,\nu'n)^{11}$C(e$^{+}\nu_{\rm e})^{11}$B.
In the O/C layer, a large fraction of $^{11}$B is also produced through the
$\nu$-process of $^{12}$C($\nu,\nu'p)^{11}$B.
However, the total mass of $^{11}$B in the O/C layer is much smaller than
in the He/C layer because the total mass of the O/C layer is only 0.1
$M_{\odot}$.
In the range of $M_r \la 4.2 M_{\odot}$ in the He/C layer, the
temperature becomes so high at the shock arrival that the produced $^7$Li and
$^{11}$B capture $^4$He to produce $^{11}$B and $^{14}$N through
$^7$Li($\alpha,\gamma)^{11}$B and $^{11}$B($\alpha,n)^{14}$N, respectively.
In the H-rich envelope, the maximum temperature does not become high enough
to allow $^3$H($\alpha,\gamma)^7$Li and
$^3$He($\alpha,\gamma)^7$Be(e$^{-},\nu_{\rm e})^7$Li, so $^7$Li is not
produced.
In this region, $^{11}$B is produced through $^{12}$C($\nu,\nu'p)^{11}$B
and the produced mass fraction is extremely small because of the small mass
fraction of $^{12}$C.

Now we describe the relation of the total ejected masses of $^7$Li and 
$^{11}$B to the parameters of the neutrino luminosity, i.e., the
total neutrino energy $E_{\nu}$ and the decay time of the 
neutrino luminosity, $\tau_{\nu}$.
Figures {\it 2a} and {\it 2b} show the ejected masses of $^7$Li and $^{11}$B 
as a function of the total energy $E_{\nu}$.
We find two clear features of the parameter dependence.
One is that the total ejected mass is almost proportional to the total
energy $E_{\nu}$ for a given decay time $\tau_{\nu}$.
The other is that the ejected mass for a given $E_{\nu}$ is insensitive to
the decay time $\tau_{\nu}$.
In Table 2 we list the sets of the two parameters adopted for the $r$-process 
calculations in the ranges of a total energy of 
$1 \times 10^{53}$ ergs $\le E_{\nu} \le 3 \times 10^{53}$ ergs and 
a decay time of 1 s $\le \tau_{\nu} \le$ 3 s.
In the given range of the total energy, the ejected masses of $^7$Li and 
$^{11}$B change within a factor of 2.7 and a factor of 2.9, respectively.
The variations due to the decay time are within 10\% for any $E_{\nu}$ values.
Hence, we can conclude that the ejected masses of $^7$Li and $^{11}$B are 
almost proportional to the total energy, and the variation is within a factor 
of 3, but insensitive to the decay time.

A simple relation between the ejected mass and the total neutrino energy 
arises from the following specific properties of the 
reaction processes that produce $^7$Li and $^{11}$B.
In the He/C-layer, $^7$Li and $^{11}$B are mainly produced through
the three reaction chains 

\begin{displaymath}
^4{\rm He}(\nu,\nu'p)^3{\rm H}(\alpha,\gamma)^7{\rm Li}(\alpha,\gamma)
^{11}{\rm B},
\end{displaymath}
\begin{displaymath}
^4{\rm He}(\nu,\nu'n)^3{\rm He}(\alpha,\gamma)^7{\rm Be(e}^-,\nu_{\rm e})
^7{\rm Li},
\end{displaymath}
\begin{displaymath}
{\rm ^{12}C(\nu,\nu'p)^{11}B. }
\end{displaymath}
At higher temperatures, a $^{11}$B($\alpha,n)^{14}$N reaction also occurs.
All these reaction sequences are triggered by neutrino spallations, and 
followed by $\alpha$-capture reactions.
The neutrino number flux is proportional to the total neutrino energy.
The spallation reaction rates for $^4$He($\nu,\nu'p)^3$H, 
$^4$He($\nu,\nu'n)^3$He, and $^{12}$C($\nu,\nu'p)^{11}$B are also proportional 
to it.
The amounts of seed nuclei of $^7$Li and $^{11}$B, namely, $^4$He and 
$^{12}$C, are determined solely by the composition of the presupernova star 
and are independent of the neutrino luminosity.
Although the mass of $^4$He and $^{12}$C in the He layer changes depending 
on the presupernova model, the ejected masses of $^7$Li and $^{11}$B are 
almost proportional to the total neutrino energy alone.
This is because the production regions of these nuclei are limited to
within the middle of the He layer.

We turn to the insensitivity of the ejected masses of $^7$Li and $^{11}$B to 
the decay time of the neutrino luminosity, as evidenced in Figure 2.
In the $^7$Li- and $^{11}$B-production regions, their mass
fractions do not strongly depend on the decay time.
This is because all neutrinos promptly pass through the region before the 
shock arrival, and $^7$Li and $^{11}$B are not effectively processed after the 
shock passes by.
In the bottom of the He/C layer ($M_r \lesssim 4.2 M_{\odot}$), where both 
$^7$Li and $^{11}$B are scarcely produced, the shock wave arrives earlier 
(about 10 s), and further $\alpha$-capture reactions on the produced 
$^7$Li and $^{11}$B proceed for several seconds after the shock arrival.
These $\alpha$-capture processes do not depend on $\tau_{\nu}$.
However, there is a slight difference, about 10\%, between the two 
cases of $\tau_{\nu}$= 1 and 3 s.
This arises from the competition between the shock arrival and the timescale 
of the production processes of $^7$Li and $^{11}$B through the $\nu$-process, 
which depends on $\tau_{\nu}$.
In the case of $\tau_{\nu}$=3 s, a fractional part of $^7$Li and $^{11}$B is 
still being produced through the $\nu$-process even after the shock arrival.
In the case of $\tau_{\nu}$=1 s, however, the $\nu$-process ends shortly 
before the shock arrival, so that the produced $^7$Li and $^{11}$B 
capture $^4$He when the shock arrives.
This difference leads to a small decrease in the amounts of $^7$Li and 
$^{11}$B in the case of $\tau_{\nu}$=1 s compared to $\tau_{\nu}$= 3 s.

Finally, we compare our result with the ejected masses of $^7$Li and
$^{11}$B in the S20A model of WW95.
In the case of $E_{\nu}= 3 \times 10^{53}$ ergs and $\tau_{\nu}$=3 s, the 
obtained masses of $^7$Li and $^{11}$B in our calculation are 
$7.46 \times 10^{-7}$ and $1.92 \times 10^{-6}$ $M_{\odot}$, 
respectively.
The masses of $^7$Li and $^{11}$B in the S20A model are 
$6.69 \times 10^{-7}$ and $1.85 \times 10^{-6} M_{\odot}$, 
respectively (see Fig. 2, {\it solid horizontal line}).
Our result is in reasonable agreement with WW95, such that 
the calculated masses of $^7$Li and $^{11}$B are 12\% and 4\% larger than the 
corresponding masses in the S20A model.

In order to investigate the effects of the different temperatures of 
$\nu_{\rm e}$ and $\bar{\nu}_{\rm e}$, we also calculated the masses of 
$^7$Li and $^{11}$B with $T_{\nu_{\rm e}}=T_{\bar{\nu}_{\rm e}}$=4 MeV/$k$, 
as adopted in WW95.
The obtained masses of $^7$Li and $^{11}$B are 9 \% and 1 \% larger than 
the corresponding masses in the S20A model, so that the difference becomes 
even smaller.
As shown in this section, the main contribution of the $\nu$-process reactions 
is neutral-current interactions of $\nu_{\mu,\tau}$ and 
$\bar{\nu}_{\mu,\tau}$ due to their temperatures being higher than those of 
$\nu_{\rm e}$ and $\bar{\nu}_{\rm e}$.
Therefore, the difference due to the temperatures of $\nu_{\rm e}$ and 
$\bar{\nu}_{\rm e}$ does not affect very much the ejected masses of $^7$Li 
and $^{11}$B.

This agreement confirms that the overproduction problem of 
$^{11}$B in the context of the GCE of light elements still remains 
independent of specific supernova models.
We discuss how to solve this problem in section 4.

\subsection{Abundances of the $r$-Process Elements}

The final total isotopic distribution of the ejected mass of the $r$-process 
elements is presented as a function of mass number in Figure \ref{final} for 
the three sets of neutrino luminosities listed in Table \ref{table1}.
The dashed line refers to the case with {\it low} total neutrino energy, 
$E_{\nu} = 1.0 \times 10^{53}$ erg, and {\it long} decay time, 
$\tau_{\nu} = 3.0$ s (LL model).  
The solid line corresponds to the case with $E_{\nu} = 1.0 \times 10^{53}$ 
ergs and {\it short} decay time, $\tau_{\nu} = 1.0$ s (LS model).  
The dotted line shows the case with {\it high} total neutrino energy, 
$E_{\nu} = 3.0 \times 10^{53}$ erg, and {\it long} decay time, 
$\tau_{\nu} =3.0$ s (HL model). 
We can observe in this figure that the third peak elements are synthesized 
in all three models, although the height of the peak and the ejected mass 
of these elements largely differ.

Let us first compare the result of the LL model with that of the LS model in 
order to investigate the dependence of the final abundance distribution on the 
decay time.  
Figure \ref{final} shows that the ejected mass in the LS model is larger 
than in the LL model. 
Since the value of $E_{\nu}$ is the same between these two models, the 
difference of $\tau_{\nu}$ leads directly to a different value for the 
neutrino luminosity (see eq. [1] and Table 2).  
In the LS model with the short decay timescale, $\tau_{\nu} = 1.0$ s, the peak 
luminosity $L_{\nu_i,0}$ is higher than in the LL model 
with the long decay time scale, $\tau_{\nu} = 3.0$ s, because of a common 
total neutrino energy, $E_{\nu} = 1.0 \times 10^{53}$ erg. 
It is known that the mass ejection rate increases as fast as the luminosity 
increases, i.e., $\dot{M} \propto L_{\nu}^{5/2}$ (Woosley et al. 1994); 
the mass ejection at high luminosity is dominant in the whole wind.
The total ejected mass is mainly determined by the peak 
luminosity, which is in turn related by the decay time.
Accordingly, a shorter decay time leads to a larger total ejected mass.

Second, we find the interesting fact that the abundance ratio of the third- 
to the second-peak elements in the LL model is larger than in the LS model.  
The reason is as follows:  
From previous studies of neutrino-driven winds, favorable conditions for a 
successful $r$-process have been identified: they are a higher 
entropy ($s/k$), shorter dynamical timescale ($\tau_{dyn}$), lower electron 
fraction ($Y_e$; e.g., Meyer \& Brown 1997), and lower asymptotic temperature 
($T_{out}$; Terasawa et al. 2002; Wanajo et al. 2002; 
Otsuki, Mathews, \& Kajino 2003).
In our present model calculations, $Y_e$ and $T_{out}$ are almost the same 
because we employ a common neutrino temperature and the same outer boundary 
conditions.
It was also found that the entropy and the dynamical time scale become 
larger as the neutrino luminosity is lower (Qian \& Woosley 1996; 
Otsuki et al. 2000; Sumiyoshi et al. 2000). 
The values of $s/k$ and $\tau_{dyn}$ change depending on $\tau_{\nu}$ 
through the change of the neutrino luminosity.
Even combining the above theoretical findings, one cannot simply explain the 
difference of the third-to-second peak abundance ratios between the LS model 
and the LL model, because the resulting effects from the change in $s/k$ 
and $\tau_{dyn}$ counteract the production efficiency of the $r$-process 
elements. 
From our numerical calculations, we find that the gain for the 
$r$-process nucleosynthesis due to the increase of entropy is quantitatively 
larger than the loss due to the increase of the dynamical timescale.  
This means that more abundant third-peak elements relative to second-peak 
elements are synthesized as the luminosity becomes lower. 
In both the LS and LL models, the efficiency of producing the 
third-peak elements is  higher for $L_{\nu_i} = L_{\nu_i,{\rm end}}$ than for  
$L_{\nu_i} = L_{\nu_i,0}$ and $L_{\nu_i, {\rm half}}$. 
In addition, as we discussed in the previous paragraph, the peak luminosity 
in the LL model is lower than in the LS model (see Table 2). 
For these reasons, the efficiency of producing the third-peak elements 
in the LL model is most prominent, leading to a larger abundance ratio of 
the third- to the second-peak elements in the LL model than in the LS model.

We next consider the dependence on the total neutrino energy $E_{\nu}$. 
 We compare the results of the LL and HL models in Figure \ref{final}. 
Since these two models have different total energies (see Table 2) with a  
common decay time, the total ejected mass is larger, and the third- to 
second peak ratio is smaller in the HL model than in the LL model for the 
reasons discussed in the previous paragraphs.

Finally, we compare the HL model with the LS model.
We obtain the result that the pattern of isotopic abundance distribution is 
exactly the same in these two models, although the ejected mass in the HL 
model is larger.
This is because $L_{\nu_i,0}, L_{\nu_i,{\rm half}}$, and 
$L_{\nu_i,{\rm end}}$, defined 
below equation (5) (see also Table 2) depend only on $E_{\nu}/\tau_{\nu}$ 
(see eq. [1]) and thus their values are exactly the same between these models.
As a result, we find that the key quantity is the neutrino luminosity for 
determining the pattern of the $r$-process abundance distribution in our study.

In summary, a lower peak luminosity is preferable 
in order to obtain a successful $r$-process abundance pattern.
In Figure \ref{final} we also display the comparison of our calculated 
results with the solar $r$-process abundance pattern 
(K\"appeler, Beer, \& Wisshak 1989) shifted to the value of the LL model 
at the second peak.  
We can conclude that the LL model is the best among our adopted three models 
in order to reproduce the solar $r$-process abundance pattern.
Note that if we change the decay time of the HL model 
($E_{\nu}= 3 \times 10^{53}$ erg) to 9 s, the pattern of 
the abundance distribution becomes the same as for the LL model.
This modification of the decay time is discussed in section 4.3.

\section{Discussion}

In this section, we discuss the constraint on the supernova neutrinos from the 
contribution of the two nucleosynthesis processes to GCE.
We then propose new supernova neutrino modeling to solve the 
overproduction problem of $^{11}$B.
We also discuss the consistency of the $r$-process abundance pattern 
using the new model.

\subsection{GCE of $^{11}$B}

Let us first discuss the overproduction problem of $^{11}$B.
Recently, the studies on the GCE of the light elements
have shown that $^7$Li and $^{11}$B originate from supernova explosions 
as well as GCR interactions with the ISM (e.g., Olive et al. 1994).
The contribution from supernova explosions is evaluated so that the
predicted $^{11}$B/$^{10}$B ratio at solar metallicity agrees with the
meteoritic ratio.
The evaluated contribution of $^{11}$B from supernovae is smaller than
that predicted from the supernova explosion models in WW95.
Several authors have investigated the GCE of the light elements and have 
introduced a reduction factor $f_\nu$ as the ratio of the amount of $^{11}$B 
determined in the GCE model to that evaluated in WW95 (e.g., Vangioni-Flam 
et al. 1996; Fields \& Olive 1999).
Namely, the amount of $^{11}$B is the same as that in WW95 for $f_{\nu} = 1$ 
and is less than that for $f_{\nu} < 1$.

Fields et al. (2000), Ramaty et al. (2000a, 2000b), and 
Alib\'es, Labay, \& Canal (2002) have evaluated the factor $f_\nu$ 
to be 0.40, 0.18, 0.28, and 0.29, respectively.
The scatter in $f_\nu$ is mainly caused by the different treatment of GCRs: 
the assumed chemical composition and energy spectra of the GCRs.
Since the factor $f_{\nu}$ still depends on the treatment of GCRs and has not 
been precisely determined, we set an acceptable range of 
the reduction factor to be 
\begin{equation}
0.18 \le f_{\nu} \le 0.40.
\end{equation}
The ejected masses of $^{11}$B with the largest and smallest values 
as denoted by $f_{\nu}=0.40$ and 0.18 together with that of WW95 
are shown in Figure {\it 2b}.

Figure {\it 2b} shows that the ejected mass of $^{11}$B meets with the above 
range of $f_{\nu}$ (see eq. [6]) only when the total neutrino energy is 
as low as $E_{\nu} \la 1.2 \times 10^{53}$ erg.
This energy is much lower than $3 \times 10^{53}$ erg, which is the value 
used in WW95 in accordance with the value constrained from the observation 
of SN 1987A.
In this energy range the ejected mass of $^7$Li is also smaller than that in
WW95 as shown in Figure {\it 2a}.

\subsection{$r$-Process Nucleosynthesis Constraint}

We turn now to the constraint from the $r$-process nucleosynthesis.
The total ejected mass of heavy $r$-process elements, $M_{eject}$, is between 
$9.0 \times 10^{-6} M_{\odot}$ (for the LL model) and 
$1.1 \times 10^{-4} M_{\odot}$ (for the HL model).
This range is consistent with the GCE of the $r$-process elements:
assuming that the Type II supernova rate is on the order of $10^{-2}$ 
yr$^{-1}$ over the entire history of Galactic evolution, the current mass 
of the $r$-process elements in the Galaxy is estimated to be 
$9 \times 10^2 M_{\odot}$ (for the LL model) and 
$1.1 \times 10^4 M_{\odot}$ (for the HL model).
Since the total baryonic mass of the Galaxy is $\sim 10^{11} M_{\odot}$, our 
models lead to a present mass fraction in $r$-process elements of the order 
of $\sim 10^{-8}$ (for the LL model) and $\sim 10^{-7}$ (for the HL model).
These values are in reasonable agreement with the observed solar mass fraction 
of $\sim 10^{-7}$.
We recall furthermore the discussion in section 3.2 that the third-to-second 
peak ratio is sensitive to the neutrino luminosity and that the LL model 
in Table 2 ($E_{\nu}=1.0 \times 10^{53}$ ergs and $\tau_{\nu}$ = 3.0 s) is 
most favorable for explaining the observed $r$-process abundance pattern.

Thus, summarizing the constraints from the two nucleosynthesis processes, 
a neutrino luminosity with $E_{\nu}=1.0 \times 10^{53}$ ergs can resolve 
the overproduction problem of $^{11}$B and the $r$-process abundance pattern 
as long as the decay $\tau_{\nu}$ is longer than or equal to 3 s.

\subsection{New Supernova Neutrino Model}

The neutrino luminosity model suggested above, however, encounters
a potential conflict between the total neutrino energy $E_{\nu}$ and the
gravitational mass of the neutron star formed in Type II supernova
explosion.
Lattimer \& Yahil (1989) suggested an approximate relation between the
gravitational binding energy $E_{\rm BE}$ and the neutron star 
mass $M_{\rm NS}$.
Since it is known that almost 99 \% of the binding energy is released as
supernova neutrinos, $E_{\rm BE}$, is equal to $E_{\nu}$ to a very good
approximation.
Their suggested relation is therefore expressed as
\begin{equation}
E_{\rm BE} \approx E_{\nu} \approx 1.5 \times 10^{53}
\left( \frac{M_{\rm NS}}{M_{\odot}} \right)^2 \quad {\rm ergs .}
\end{equation}
This formulation is shown to be a reasonable approximation in the theoretical
studies of several non-relativistic potential models and field theoretical
models (Prakash et al. 1997; Lattimer \& Prakash 2001).
Using this formulation, the total neutrino energy for a neutron star
mass of 1.4 $M_{\odot}$ turns out to be
\begin{equation}
2.4 \times 10^{53} \  {\rm erg} \  \la E_{\nu}
\la \  3.5 \times 10^{53} \  {\rm erg,}
\end{equation}
within $\pm 20$ \% error bars.
This range is displayed by two vertical lines in Figures {\it 2a} and {\it 2b}.
Although we summarized above that the most suitable total neutrino energy is
$E_{\nu} = 1.0 \times 10^{53}$ ergs from the constraints on the two
nucleosynthesis processes, it is inconsistent with equation (8) which is
based on the neutron star formation conjecture in a Type II supernova 
explosion.

In order to solve this inconsistency, we modify the temperature of 
$\nu_{\mu,\tau}$ and $\bar{\nu}_{\mu,\tau}$,
which we set to be 8 MeV/$k$ as in WW95, and reset the decay time of the 
neutrino luminosity.
Rauscher et al. (2002) have tried to reduce the amount of $^{11}$B
by decreasing these neutrino temperatures from $8$ MeV/$k$ to $6$ MeV/$k$.
Although the spectrum of the neutrinos emitted from a proto-neutron star
has not been determined, some recent studies show the neutrino temperature
to be smaller than 8 MeV/$k$ (e.g., Myra \& Burrows 1990; Keil et al. 2003).
We therefore adopt the same lower temperature for $\nu_{\mu,\tau}$ and 
$\bar{\nu}_{\mu,\tau}$ of 6 MeV/$k$.

In section 3.2 we showed that the LL model is the most favorable for 
explaining the observed $r$-process abundance pattern and that a lower peak 
luminosity is preferable.
However, the total neutrino energy of the LL model is outside the range of 
equation (8).
In order to cure the situation, we set a longer decay time for the neutrino 
luminosity: $\tau_{\nu}$= 9 s.
Using the long decay time, we can attain the total neutrino energy in the 
range of equation (8) while still preserving the lower peak neutrino 
luminosity (see eq. [1]).

We show the ejected masses of $^7$Li and $^{11}$B in the case of 
$T_{\nu_{\mu,\tau}}=T_{\bar{\nu}_{\mu,\tau}}$= 6 MeV/$k$ and 
$\tau_{\nu}$= 9 s by a dashed line in Figure 2.
The ejected masses of $^7$Li and $^{11}$B decrease drastically compared to
those in the case of $T_{\nu_{\mu,\tau}}=T_{\bar{\nu}_{\mu,\tau}}$=8 MeV/$k$.
This is because smaller amounts of seed nuclei, such as $^3$H and $^3$He for
the production of $^7$Li and $^{11}$B, are provided from the neutrino
spallation of $^4$He due to smaller cross sections of neutral-current
interactions at the lower neutrino temperature.
We also consider the case of a decay time of 3 s (Fig. 2, {\it dot-dashed 
line}).
As shown in section 3.1, the masses of $^7$Li and $^{11}$B scarcely depend on 
the decay time even in the case of
$T_{\nu_{\mu,\tau}}=T_{\bar{\nu}_{\mu,\tau}}$= 6 MeV/$k$.

Let us compare the ejected mass of $^{11}$B again with that required from 
the GCE models.
It now turns out to be in the proper range required from the GCE model 
analyses (eq. [6]), where the total neutrino energy $E_{\nu}$ is between 
$1.5 \times 10^{53}$ and $3.4 \times 10^{53}$ ergs (see Fig. 2{\rm b}).
The proper range clearly overlaps with the range deduced from the restriction
with the neutron star mass constraint in equation (8).
Hence, we can conclude that the ejected mass of $^{11}$B required from the GCE
models is successfully reproduced with the appropriate total neutrino energy
when one adopts the neutrino temperature of 6 MeV/$k$ and a decay time 
of the neutrino luminosity of 9 s.
The ejected masses of $^7$Li and $^{11}$B are
\begin{displaymath}
 2.3 \times 10^{-7} M_{\odot} \le M(^{7}{\rm Li}) \le
3.1 \times 10^{-7} M_{\odot},
\end{displaymath}
\begin{displaymath} 
5.2 \times 10^{-7} M_{\odot} \le M(^{11}{\rm B}) \le
7.4{\times}10^{-7} M_{\odot}, 
\end{displaymath} 
respectively.

Now we consider the effects of these lower neutrino temperatures on
the $r$-process nucleosynthesis.
We adopt the MLL ({\it modified} LL) model in Table 2 with a total 
neutrino energy of $3 \times 10^{53}$ ergs and a decay time of the neutrino 
luminosity of $9$ s.
The neutrino luminosities of this model are the same as those of the 
LL model (see Table 2).
When we reduce the neutrino temperature from $8$ MeV/$k$ to $6$ MeV/$k$,
the dynamical timescale of expansion in the neutrino-driven winds becomes 
slightly longer, by only a few milliseconds.
However, this effect does not drastically change the $r$-process abundance 
pattern, although the ejected mass is different, mainly owing to the different 
$t_{end}$ values (see eq. [5]). 
Figure 4 shows the resulting abundance pattern of the MLL model
($T_{\nu_{\mu,\tau}}=T_{\bar{\nu}_{\mu,\tau}} = 6$ MeV/$k$), with stars
showing the solar $r$-process abundances (K\"appeler et al. 1989) shifted to 
the value of the MLL model at the second peak.
By comparison with the LL model (the dashed line, which is the same as that 
in Figure 3), which is the best case at 
$T_{\nu_{\mu,\tau}}=T_{\bar{\nu}_{\mu,\tau}} = 8$ MeV/$k$, 
the third-to-second peak ratio slightly increases.
We therefore conclude that a lower neutrino temperature is preferable for 
not only light elements but also heavy $r$-process elements.
These low temperatures are acceptable from the point of view of recent 
studies on supernova neutrinos (Myra \& Burrows 1990; Keil et al. 2003).

Finally, we obtained the mass of $^{11}$B suitable for the GCE of
the light elements and a successful $r$-process abundance pattern
satisfying the restriction on the total neutrino energy from a typical
neutron star mass when we adopt the temperature of $\nu_{\mu,\tau}$ and 
$\bar{\nu}_{\mu,\tau}$ to be 6 MeV/$k$ and
the decay time of the neutrino luminosity to be 9 s.

\section{Summary}
We investigated the influence of the total neutrino energy and the decay 
time of the neutrino luminosity on the light-element synthesis.
We also investigated the $r$-process nucleosynthesis in neutrino-driven winds
with the same neutrino luminosity.
We summarize our findings.

First, the ejected masses of $^7$Li and $^{11}$B are roughly proportional to
the total neutrino energy.
The difference due to the decay time of the neutrino luminosity is
small, i.e., within 20\%.
Therefore, in order to obtain the ejected masses of the light elements 
precisely, it is important to determine the total neutrino energy 
rather than the decay time of the neutrino luminosity.

Second, results of the $r$-process nucleosynthesis depend on the peak 
neutrino luminosity, which depends on $E_{\nu}$/$\tau_{\nu}$.
Although the $r$-process is sensitive to the total energy of the neutrinos,
only the ejected mass is affected strongly.
In order to obtain a successful $r$-process abundance pattern, a low peak 
luminosity is preferable, such as obtained in the LL model 
(see Table \ref{table1}).

We discussed the contributions of $^{11}$B and $r$-process elements from
supernovae to the Galactic chemical evolution.
We found that the preferred total neutrino energy is about 
$1.0 \times 10^{53}$ ergs and the decay time of the neutrino flux should be 
longer than or equal to 3 s when we set the temperature of $\nu_{\mu,\tau}$ 
and $\bar{\nu}_{\mu,\tau}$ to be 8 MeV/$k$.

However, assuming the mass of the proto-neutron star formed in a supernova 
explosion to be 1.4 $M_{\odot}$, the total neutrino energy is evaluated to be 
about $3 \times 10^{53}$ ergs (Lattimer \& Prakash 2001).  
The model mentioned above is inconsistent with this total energy.  
We propose a new supernova neutrino modeling to overcome this inconsistency: 
to reduce the temperature of $\nu_{\mu,\tau}$ and $\bar{\nu}_{\mu,\tau}$
to $6$ MeV/$k$, as used in Rauscher et al. (2002) 
and to raise the decay time of the neutrino luminosity to 9 s.
With these modifications of the supernova neutrino modeling we successfully 
obtain the proper ejected mass of $^{11}$B and the $r$-process abundance pattern.

\acknowledgments

We would like to thank Koichi Iwamoto, Ken'ichi Nomoto, and Toshikazu 
Shigeyama for providing the data for the internal structure of progenitor 
model 14E1 and for helpful discussions.
We are grateful to Shoichi Yamada for collaboration on the hydrodynamic 
simulations of neutrino-driven winds based on his original 
numerical code for hydrodynamics.
We are also indebted to Nobuyuki Iwamoto, Kaori Otsuki, Grant J. Mathews, 
and Takeru Suzuki for their many valuable discussions.
T. Y. and M. T. are supported by Research Fellowships of the Japan Society 
for the Promotion of Science for Young Scientists (12000289 and 13006006).
This work has also been supported in part by Grants-in-Aid for Scientific 
Research (12047233, 13640313, 13740165, 15740160) and 
for Specially Promoted Research (13002001) of the Japan Society 
for Promotion of Science and the Ministry of 
Education, Science, Sports and Culture of Japan.



%\appendix

%\section{Appendicial material}

\clearpage


\begin{figure}
\plotone{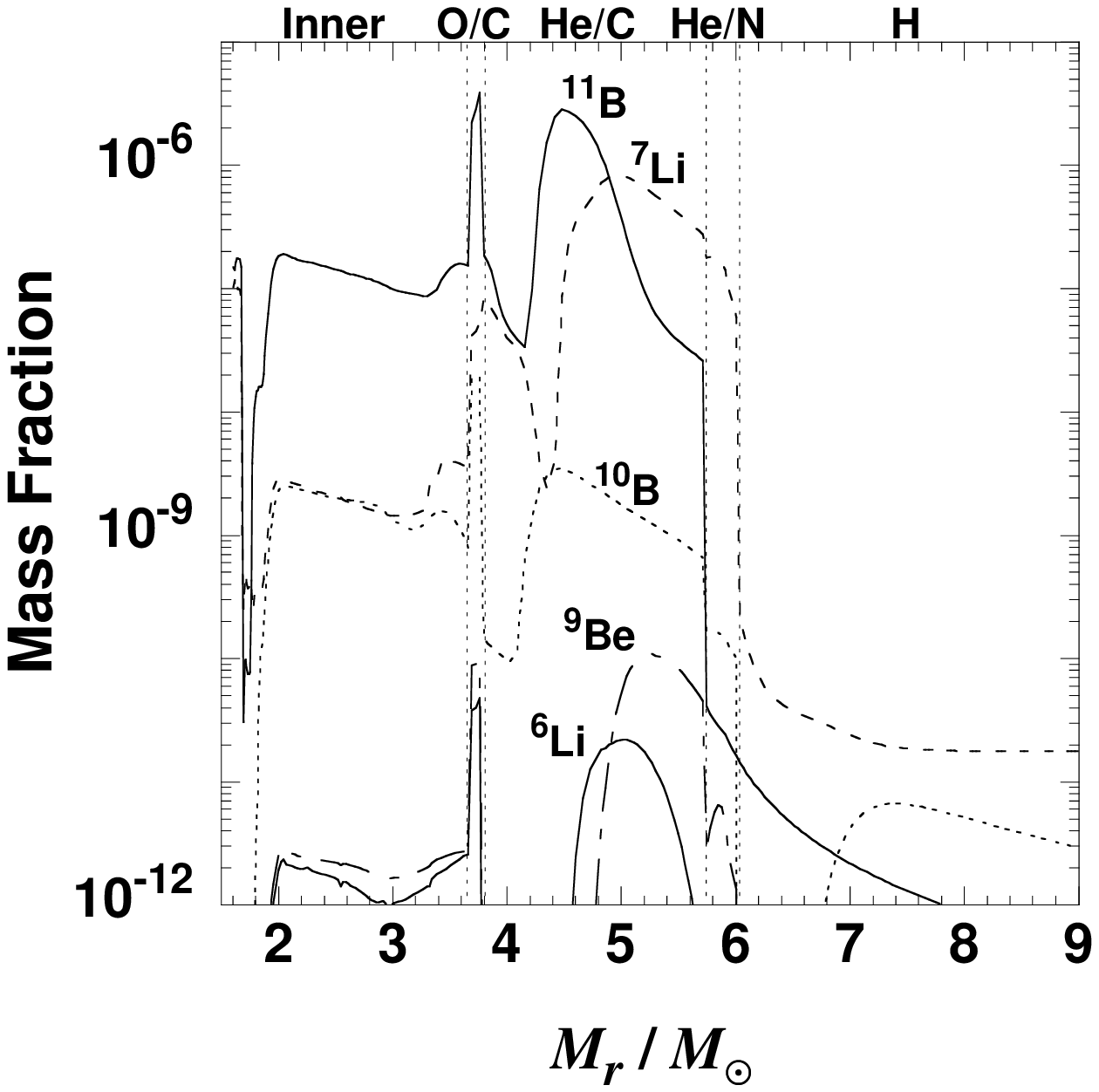}
\caption{Distribution of mass fractions of the light elements (Li, Be, 
and B) in the case of $E_{\nu}=3 \times 10^{53}$ ergs and $\tau_{\nu}$=3 s.
The mass fraction of $^7$Li means the sum of the mass fractions of $^7$Li and 
$^7$Be.
The mass fraction of $^{11}$B means the sum of the mass fractions of 
$^{11}$B and $^{11}$C.
The inner layers, the O/C layer, the He/C layer, the He/N layer, and 
the H-rich envelope (as denoted by ^^ ^^ Inner'', ^^ ^^ O/C'', ^^ ^^ He/C'', 
^^ ^^ He/N'', and ^^ ^^ H'' in the top panel) correspond to the ranges of 
the mass coordinate, 
$M_r \le 3.7 M_{\odot}$, $3.7 M_{\odot} \le M_r \le 3.8 M_{\odot}$, 
$3.8 M_{\odot} \le M_r \le 5.8 M_{\odot}$, 
$5.8 M_{\odot} \le M_r \le 6.0 M_{\odot}$, and $M_r \ge 6.0 M_{\odot}$, 
respectively.
}
\end{figure}

\clearpage 

\begin{figure}
\plotone{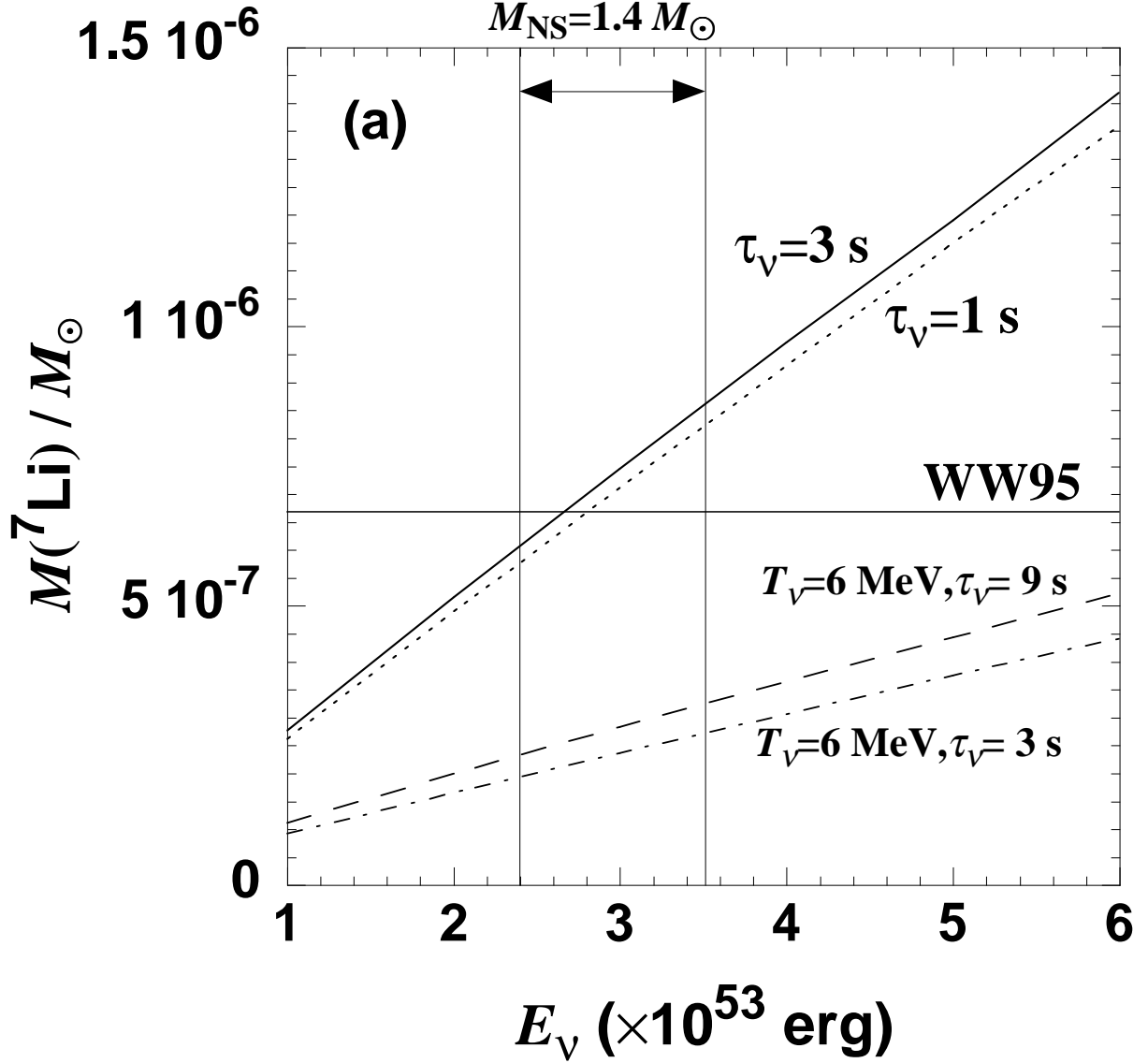}
\caption{Ejected masses of $^7$Li ({\rm a}) and $^{11}$B ({\rm b}) as a 
function of the total neutrino energy $E_{\nu}$.
The solid and dashed lines show the ejected masses in the cases of decay 
times of the neutrino luminosity of $\tau_{\nu}$=3 and 1 s, respectively.
The solid horizontal line denoted by ^^ ^^ WW95'' shows the ejected mass 
given by the S20A model in WW95.
The range between solid vertical lines denotes the total energy range of 
supernova neutrinos determined from the binding energy of a 1.4
 $M_{\odot}$ neutron star.
The solid horizontal lines in ({\rm b}) show the evaluated $^{11}$B masses 
in the cases of $f_\nu$= 0.40 and 0.18, where $f_{\nu}$ is the reduction 
factor introduced from the discussion of the GCE of the light elements.
Dashed and dot-dashed lines show the ejected masses in the cases of the low 
temperature of $\nu_{\mu,\tau}$ and  $\bar{\nu}_{\mu,\tau}$ of 
6 MeV/$k$ for decay times of 9 and 3 s, respectively (see section 4.3).
See the text for details.
}
\end{figure}

\clearpage

\begin{figure}
\figurenum{2}
\plotone{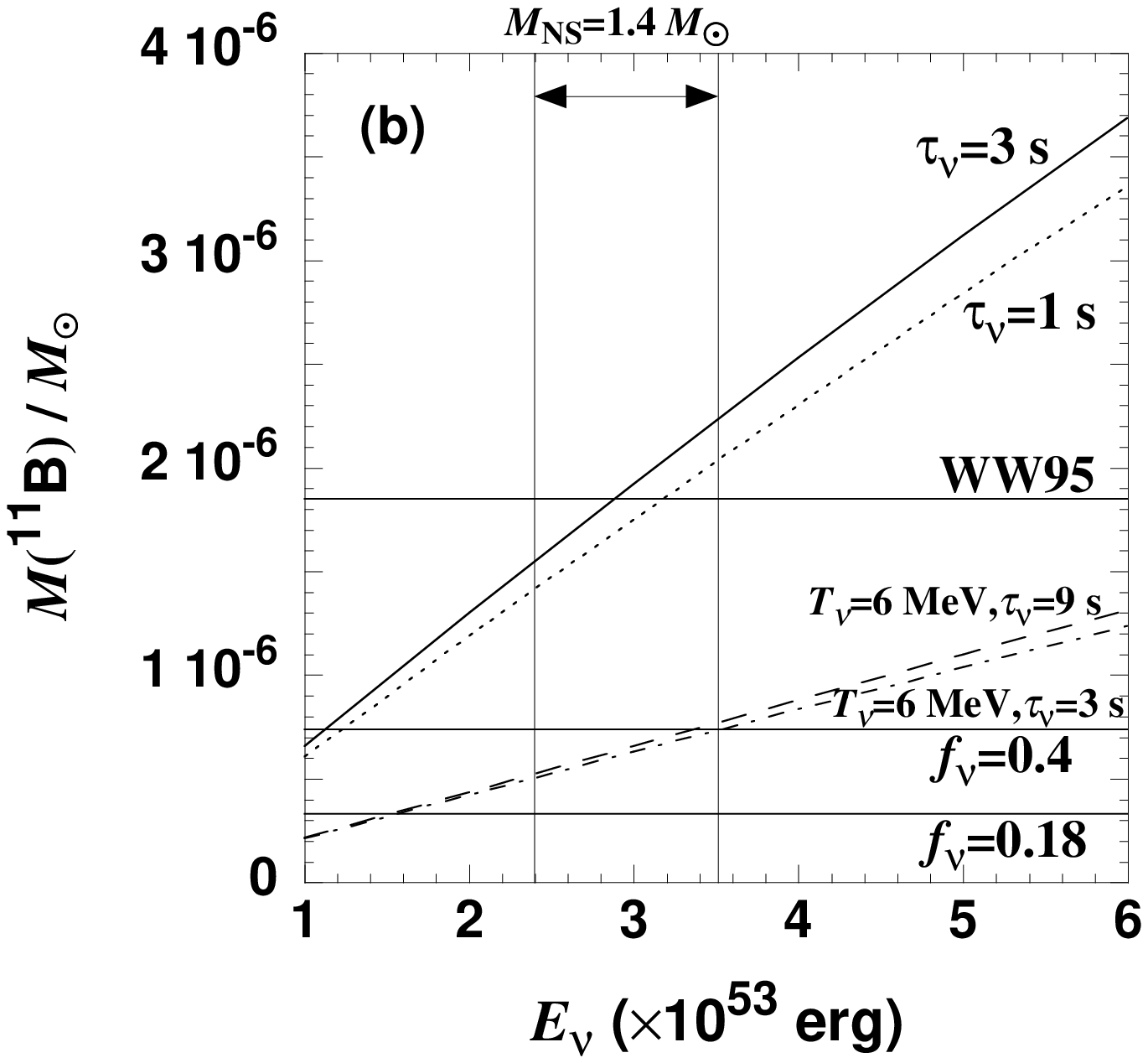}
\caption{
{\it Continued}.
}
\end{figure}

\clearpage

\begin{figure}
\plotone{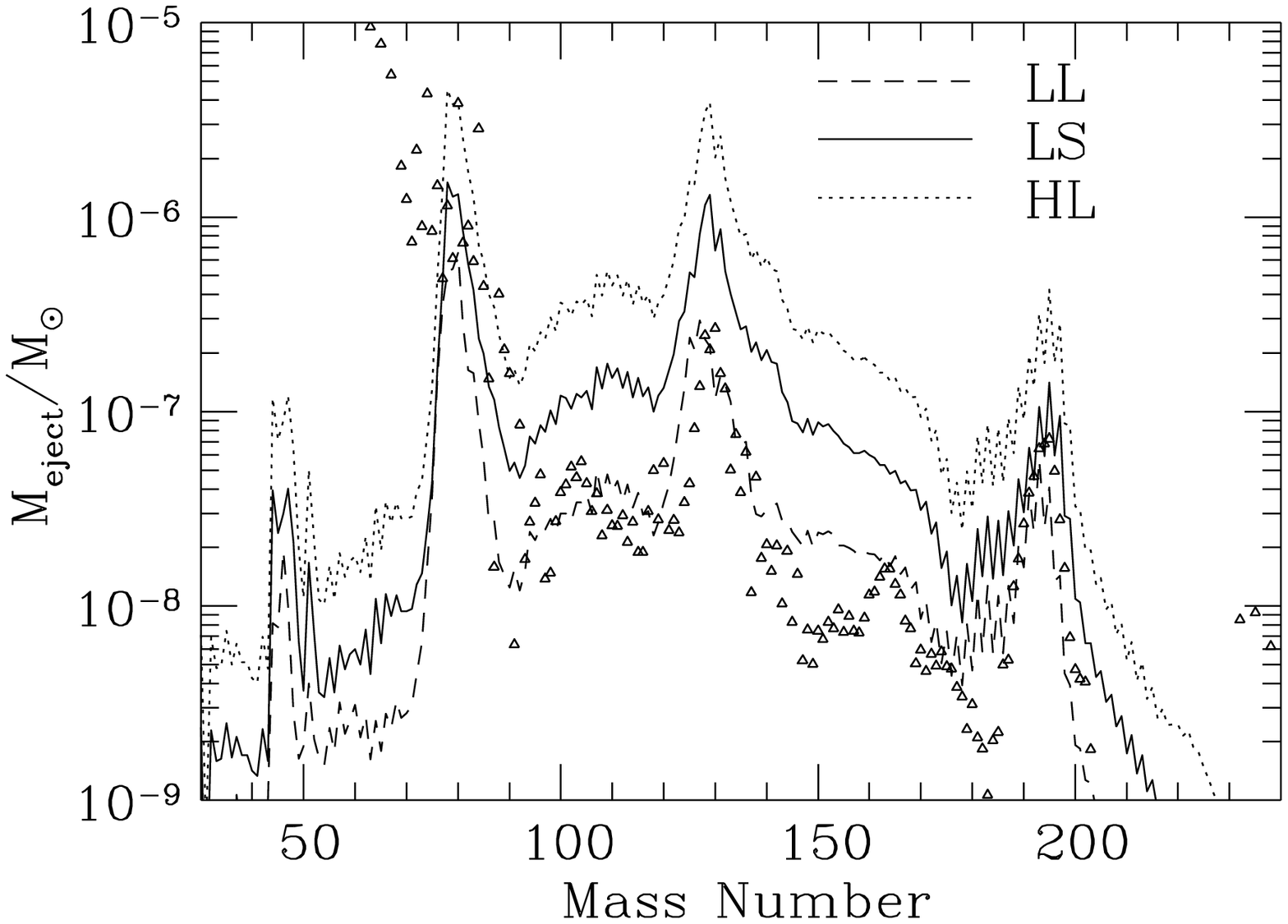}
\caption{Final total isotopic distribution of ejected mass as a function 
of mass number. 
The dashed, solid, and dotted lines refer to the LL, LS, and HL models, 
respectively (see Table 2). 
Triangles also show the observed solar $r$-process abundances
(K\"appeler et al. 1989) shifted to the value of the LL model at the 
second peak.\label{final}
}
\end{figure}

\clearpage

\begin{figure}
\plotone{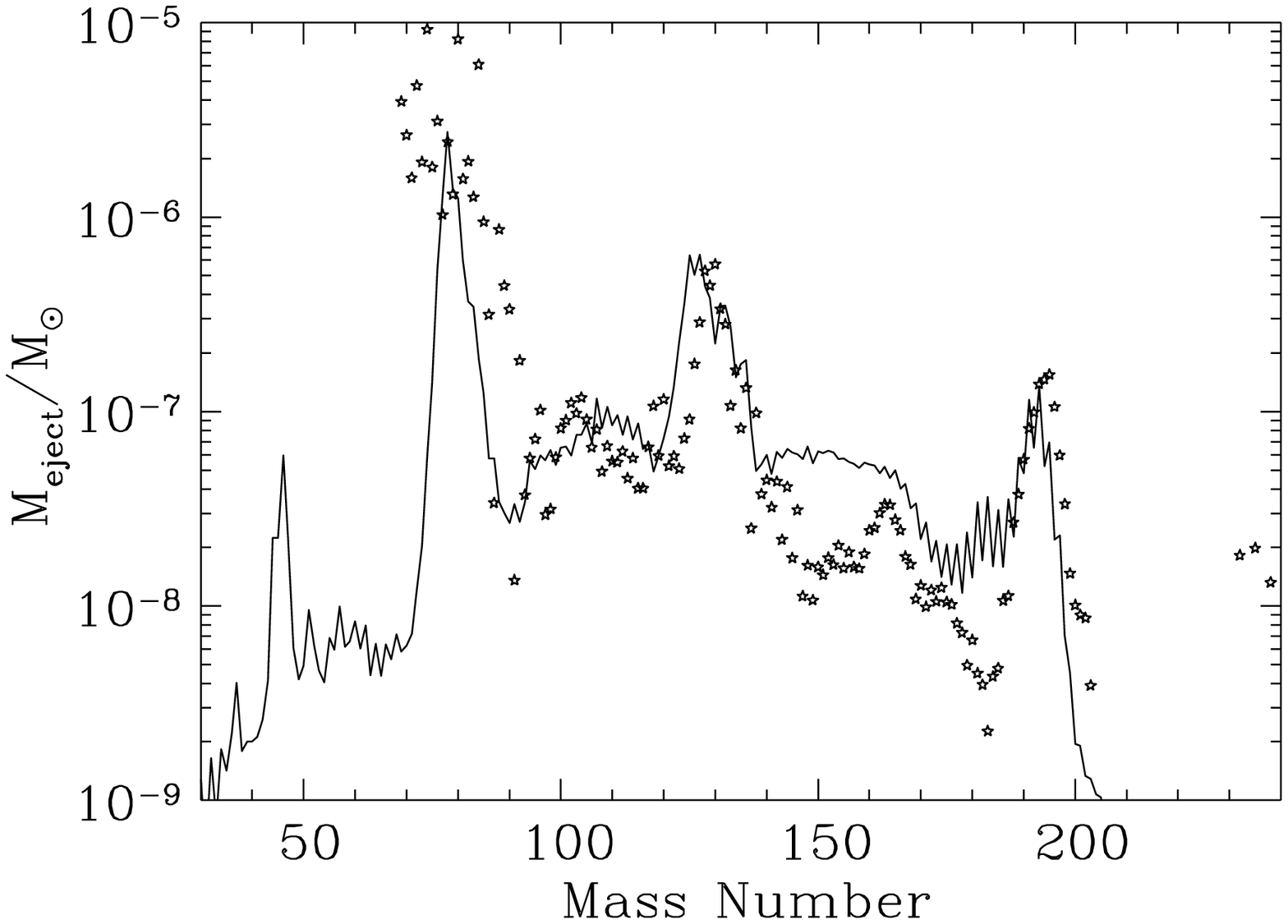}
\caption{Final total isotopic distribution of ejected mass as a function 
of mass number of the MLL model (see Table 2).
Stars show the observed solar $r$-process abundances
(K\"appeler et al. 1989) shifted to the value of the MLL model 
at the second peak.\label{f4}
}
\end{figure}

%% If you are not including electonic art with your submission, you may
%% mark up your captions using the \figcaption command. See the 
%% No more than seven \figcaption commands are allowed per page, 
%% so if you have more than seven captions, insert a \clearpage 

%% Tables should be submitted one per page, so put a \clearpage before

%% Two options are available to the author for producing tables:  the
%% deluxetable environment provided by the AASTeX package or the LaTeX
%% table environment.  Use of deluxetable is preferred.
%%

%% Three table samples follow, two marked up in the deluxetable environment,
%% one marked up as a LaTeX table.

%% In this first example, note that the \tabletypesize{}
%% command has been used to reduce the font size of the table.
%% Note also that the \label command needs to be placed 

\clearpage

\begin{deluxetable}{cccccccccc}
\tabletypesize{\scriptsize}
\tablecaption{Nuclear Reaction Network Used for Light-Element 
Nucleosynthesis}
\tablewidth{0pt}
\tablehead{
Element & $A$ & Element & $A$ & Element & $A$ & Element & $A$ & Element & $A$
}
\startdata
n  & 1       & N  & 12-17 & Si & 25-33 & Sc & 40-50 & Ni & 54-67 \\
H  & 1-3     & O  & 14-20 & P  & 27-35 & Ti & 42-52 & Cu & 57-69 \\
He & 3,4,6   & F  & 17-21 & S  & 29-38 & V  & 44-54 & Zn & 59-72 \\
Li & 6-9     & Ne & 18-25 & Cl & 31-40 & Cr & 46-56 & Ga & 61-74 \\
Be & 7,9-11  & Na & 20-26 & Ar & 33-44 & Mn & 48-58 & Ge & 68-74 \\
B  & 8,10-12 & Mg & 21-28 & K  & 35-46 & Fe & 50-62 &    & \\
C  & 11-15   & Al & 23-30\tablenotemark{a} & Ca & 37-49 & Co & 52-63 &    & \\
\enddata
\tablecomments{Here $A$ is the mass number.}
\tablenotetext{a}{The ground state and the isomeric state of $^{26}$Al are 
treated as separate species.}
\end{deluxetable}

\clearpage

\begin{deluxetable}{ccccccc}
\tabletypesize{\small}
\tablecaption{The Adopted Parameter Sets of Neutrino Luminosity Models.
\label{table1}}
\tablewidth{0pt}
\tablehead{
\colhead{Model} & 
\colhead{$E_{\nu}$ (erg)} & \colhead{$\tau_{\nu}$ (s)} & 
\colhead{$L_{\nu_i, 0}$ (ergs s$^{-1}$)} &
\colhead{$L_{\nu_i, {\rm half}}$ (ergs s$^{-1}$)} & 
\colhead{$L_{\nu_i, {\rm end}}$ (ergs s$^{-1}$)} & 
\colhead{$t_{{\rm end}}$ (s)}
}
\startdata
\cutinhead{$T_{\nu_{\mu,\tau}}=T_{\bar{\nu}_{\mu,\tau}}=8.0$ MeV/$k$}
LL  & $1.0 \times 10^{53}$ & $3.0$ & $5.56 \times 10^{51}$ &
$4.42 \times 10^{51}$ & $3.50 \times 10^{51}$ & $1.39$\\
LS  & $1.0 \times 10^{53}$ & $1.0$ & $16.67 \times 10^{51}$ & $7.64
\times 10^{51}$ & $3.50 \times 10^{51}$ & $1.56$\\
HL  & $3.0 \times 10^{53}$ & $3.0$ & $16.67 \times 10^{51}$ & $7.64
\times 10^{51}$ & $3.50 \times 10^{51}$ & $4.68$\\
\cutinhead{$T_{\nu_{\mu,\tau}}=T_{\bar{\nu}_{\mu,\tau}}=6.0$ MeV/$k$} 
MLL & $3.0 \times 10^{53}$ & $9.0$ & $5.56 \times 10^{51}$ &
$4.42 \times 10^{51}$ & $3.50 \times 10^{51}$ & $4.16$\\
\enddata

\tablecomments{
Here $T_{\nu_e}=3.2$ MeV/$k$, $T_{\bar{\nu}_e}=5.0$ MeV/$k$, 
$E_{\nu}$ is the total neutrino energy, 
$L_{\nu_i}$ is the neutrino luminosity 
($\nu_i = \nu_e, \nu_{\mu}, \nu_{\tau}$, and their antineutrinos), 
and $t_{end}$ is the time when the values of $L_{\nu i}$
become $3.5 \times 10^{51}$ ergs s$^{-1}$.
We use the same $L_{\nu_i}$ for all neutrino species.
See the text for details.}

\end{deluxetable}

%% The following command ends your manuscript. LaTeX will ignore any text

\end{document}